%% file: neurips_2026.tex
\documentclass{article}

\usepackage[main,final,nonatbib]{neurips_2026}

\usepackage[utf8]{inputenc}
\usepackage[T1]{fontenc}
\usepackage[hidelinks]{hyperref}
\usepackage{url}
\usepackage{booktabs}
\usepackage{amsfonts}
\usepackage{amsmath}
\usepackage{nicefrac}
\usepackage{microtype}
\usepackage{xcolor}
\usepackage{colortbl}
\usepackage{graphicx}
\usepackage{listings}
\usepackage{multirow}
\usepackage{placeins}
\usepackage{pgfplots}
\pgfplotsset{compat=1.18}
\usepgfplotslibrary{groupplots}
\usepackage{capt-of}
\hypersetup{
  pdftitle={Reward-Weighted On-Policy Distillation with an Open Property-Equivalence Verifier for NL-to-SVA Generation},
  pdfauthor={Qingyun Zou, Yingze Li, Tianen Liu, Bingsheng He, Weng-Fai Wong}
}

\definecolor{codecolor}{rgb}{0.95,0.95,0.95}
\title{Reward-Weighted On-Policy Distillation with an Open Property-Equivalence Verifier for NL-to-SVA Generation}

\author{%
  Qingyun Zou \\
  National University of Singapore \\
  \href{mailto:qingyunzou@u.nus.edu}{qingyunzou@u.nus.edu}
  \And
  Yingze Li \\
  Harbin Institute of Technology \\
  \href{mailto:23b903046@stu.hit.edu.cn}{23b903046@stu.hit.edu.cn}
  \And
  Tianen Liu \\
  Southeast University \\
  \href{mailto:toliutianen@seu.edu.cn}{toliutianen@seu.edu.cn}
  \AND
  Bingsheng He \\
  National University of Singapore \\
  \href{mailto:dcsheb@nus.edu.sg}{dcsheb@nus.edu.sg}
  \And
  Weng-Fai Wong \\
  National University of Singapore \\
  \href{mailto:wongwf@nus.edu.sg}{wongwf@nus.edu.sg}
}

\begin{document}

\maketitle

\begin{abstract}
Specialized language models for translating natural language (NL) into SystemVerilog assertions (SVA) reach approximately 76\% accuracy on NL2SVA-Human, but aggregate scores obscure errors on bounded-delay and liveness specifications. Models often reuse a small set of implication templates where different temporal behavior is required. Supervised fine-tuning rewards agreement with reference tokens, whereas SVA correctness depends on property equivalence. We propose \emph{reward-weighted on-policy distillation} (RWOPD), which uses an open SymbiYosys+Z3 property-equivalence checker to select student rollouts and weight their distillation loss. A frozen 14B teacher supplies forward Kullback--Leibler (KL) supervision at each response token; equivalent rollouts receive greater weight than one-sided implications. Starting from curriculum supervised fine-tuning (SFT), RWOPD distills CodeV-SVA-14B into a Qwen2.5-Coder-7B-Instruct student. The student has the highest reported pass@1, pass@5, and pass@10 on NL2SVA-Human and NL2SVA-Machine among the compared models, including specialized baselines and 671B general-purpose models.
\end{abstract}

\section{Introduction}
\label{sec:intro}

SystemVerilog assertions (SVAs) express safety, liveness, and bounded-time obligations in formal verification of register-transfer-level (RTL) designs \cite{mehta2020sva, clarke1997model}. Because hand-writing SVAs from natural-language (NL) specifications is slow and error-prone, recent work has trained large language models to translate natural-language specifications into assertions \cite{yan2025assertllm, wu2026qimeng, kang2025fveval}. The strongest specialized model, CodeV-SVA-14B \cite{wu2026qimeng}, reports roughly 76\% functional accuracy on NL2SVA-Human. This aggregate score does not show which temporal specifications remain difficult.

The aggregate score hides differences in temporal structure. We use a three-class \emph{temporal complexity level} taxonomy: \emph{combinational} assertions contain no temporal operators, \emph{bounded temporal} assertions use implication or fixed or ranged delays, and \emph{liveness} assertions use unbounded liveness operators. An assertion belongs to the highest-complexity class of its operators. This taxonomy describes syntax; property equivalence determines correctness. Our analysis finds that both the 14B specialized model and 7B baselines overproduce a small set of implication templates (\S\ref{sec:gap}). We call this \emph{template collapse}: the generated assertion defaults to a familiar temporal structure even when the specification calls for a different one.

The training objective used by existing NL-to-SVA systems does not directly optimize that target. Supervised fine-tuning (SFT) over NL/SVA pairs uses cross-entropy to reward token-level agreement with one reference assertion, even though SVA correctness is semantic: syntactically different assertions can be equivalent, while a change to one temporal operator can change the property. This motivates evaluating the generated property and training on the model's own rollouts.

Reinforcement learning with verifiable rewards (RLVR) uses a verifier's outcome as a training signal \cite{cobbe2021verifiers,lightman2023letsverify,guo2025deepseek}. For NL-to-SVA, that outcome must compare the generated property with the intended reference. We build an open SymbiYosys+Z3 property-equivalence checker to supply this signal during training, while retaining the benchmark's Cadence JasperGold protocol for final evaluation. The checker returns equivalence, a one-sided implication, rejection, or abstention. These verdicts support semantic selection, but a scalar reward alone gives limited information about which tokens to change.

We combine this rollout-level feedback with on-policy distillation (OPD), using token-level supervision and student-sampled rollouts \cite{agarwal2024onpolicy}. Our \emph{reward-weighted on-policy distillation} (RWOPD) retains equivalent and one-sided-implication rollouts and gives equivalent rollouts greater relative weight. A frozen CodeV-SVA-14B teacher conditions on the reference assertion, while the student conditions on its own rollout; forward Kullback--Leibler (KL) divergence transfers the teacher's response-token distributions to the student. Keeping one-sided implications supplies training examples when exact equivalence is rare; weighting favors equivalent examples. A curriculum-SFT seed supplies initial coverage of all three temporal classes. Rejection-sampling supervised fine-tuning (RS-SFT) and group relative policy optimization (GRPO) are separate post-training controls initialized from that seed; the RWOPD pipeline consists of curriculum SFT followed by reward-weighted distillation.

The contributions of this paper are:

(1) \emph{Reward-weighted on-policy distillation}: an OPD method that selects and weights the teacher's forward-KL supervision using an open property-equivalence checker (\S\ref{sec:opd}).

(2) An open SymbiYosys+Z3 property-equivalence checker that supplies a license-free, second-scale verifier signal for training-time filtering and reward experiments. The checker evaluates the supported combinational and bounded temporal fragments and abstains on unsupported liveness cases; JasperGold is retained for benchmark evaluation (\S\ref{sec:pec}).\footnote{Code for the checker, RWOPD, and curriculum SFT: \url{https://github.com/zanehpn/LLM4SVA}.}

(3) \emph{Empirical results.} The 7B student trained with low-rank adaptation (LoRA) has the highest reported pass@1, pass@5, and pass@10 on both NL2SVA benchmarks among the compared models, including CodeV-SVA-14B and 671B general-purpose baselines (Table~\ref{tab:main}). On NL2SVA-Machine, the absolute gains over the teacher are 7.5\% on bounded temporal assertions and 11.4\% on liveness assertions, while accuracy on combinational assertions is unchanged (Figure~\ref{fig:per_class_machine}).

\section{Related work}
\label{sec:related}

\paragraph{Hardware generation and verifier feedback.} Code language models~\cite{chen2021codex,roziere2023code,li2023starcoder} support hardware generation at several levels. Repository-level benchmarks cover agentic bug repair~\cite{zou2026phoenix} and multilingual hardware code completion~\cite{zou2026mhrc}. For high-level synthesis (HLS), HLStrans~\cite{zou2025hlstrans} provides paired C and HLS programs with testbenches and synthesis-derived latency and resource annotations, while HLS-Seek~\cite{zou2026hlsseek} uses comparative proxy rewards and selective synthesis calls to optimize latency and resource utilization. For RTL generation, EARL~\cite{shi2026earl} uses verifiable rewards and entropy-guided policy updates that concentrate gradients on high-entropy tokens. For NL-to-SVA, prior systems structure specifications, connect verification objectives to RTL signals, and refine candidates through search~\cite{yan2025assertllm,wu2025spec2assertion,lyu2025assertgen,gupta2025sangam}. The VERT dataset augments RTL/assertion pairs~\cite{menon2025vert}, and CodeV-SVA uses RTL-grounded synthesis and bidirectional translation to construct specialized training data~\cite{wu2026qimeng}. We use CodeV-SVA's training split and frozen 14B model for student post-training. The formal-verification evaluation suite FVEval~\cite{kang2025fveval} supplies the NL2SVA benchmarks and reference-property checks used in our JasperGold evaluation; AssertionBench~\cite{pulavarthi2025assertionbench} evaluates assertion generation on curated hardware designs with verified references. AssertFix~\cite{lyu2025assertfix} uses failure diagnosis to repair assertions at inference time. Our checker supplies training-time equivalence and directional-implication verdicts for student rollouts, with abstention outside its supported bounded fragment.

\paragraph{Distillation and student adaptation.} Knowledge distillation transfers teacher predictions to a smaller student~\cite{hinton2015distilling}; OPD addresses training--inference prefix mismatch by querying the teacher on student-generated trajectories~\cite{agarwal2024onpolicy}. Variants construct a feedback-conditioned self-teacher~\cite{hubotter2026sdpo} or mix teacher and student turns while reducing teacher intervention~\cite{li2026guidedopd}. Other approaches combine distillation with reinforcement learning using scalar verifier rewards, such as those used by GRPO~\cite{shao2024deepseekmath,guo2025deepseek}: joint objectives use reverse-KL regularization~\cite{xu2025kdrl} or importance-weighted teacher guidance~\cite{wang2026distilledrl}, while correctness-gated methods modulate sampled-token rewards or teacher logits using verifier feedback~\cite{lin2026opdvr,akhondzadeh2026rgopd}. Sequential Beats Joint~\cite{li2026sequential} separates OPD and RLVR into successive stages. RWOPD selects and weights a frozen teacher's forward-KL losses using equivalence and directional-implication verdicts, with abstention on unsupported properties. Our controls compare unfiltered OPD, equivalence-only OPD, and verifier-only training; broader comparisons with joint objectives require matched implementations. Knowledge retention also matters during adaptation: Xia et al.~\cite{xia2026streaming} survey continual adaptation in streaming-data modeling, and Lin et al.~\cite{lin2026ortcl} use orthogonal rotations to mitigate forgetting in time-series foundation models. Our curriculum uses replay to retain earlier SVA classes across fixed training stages (\S\ref{sec:curriculum}).

\section{The gap between generated and reference temporal structure}
\label{sec:gap}

To examine temporal structure beyond aggregate pass@$k$, we classify references and generated SVAs by their highest-complexity operator using this taxonomy (rules and validation in Appendix~\ref{app:tcl}). NL2SVA-Human contains 79 tasks: 62 combinational, 6 bounded temporal, and 11 liveness references. CodeV-SVA-14B~\cite{wu2026qimeng} matches the reference class on 18/62 combinational, 3/6 bounded temporal, and 4/11 liveness tasks; the unfine-tuned Qwen2.5-Coder-7B-Instruct matches on 19/62, 0/6, and 4/11, respectively. In that class order, their reported generation counts are $24/46/9$ and $31/41/7$, compared with $62/6/11$ references. The excess of bounded temporal outputs is consistent with \emph{template collapse}: models reuse familiar implication forms such as \texttt{|->} across specifications with different temporal requirements.

Agreement on temporal class measures structural similarity, while semantic correctness requires property equivalence. For example, under the same clock and reset conditions, \texttt{req |-> ack} requires same-cycle acknowledgement, whereas \texttt{req |-> \#\#[1:3] ack} requires acknowledgement one to three cycles later; both belong to the bounded temporal class~\cite{mehta2020sva}. Equivalent properties can also use different operator forms, so agreement on temporal class is neither sufficient nor necessary for correctness. We therefore measure correctness with JasperGold, grouped by reference class in Figure~\ref{fig:per_class_machine}. For training, curriculum SFT and temporal-token weighting provide operator coverage and emphasize timing and liveness, while RWOPD uses property-equivalence verdicts to select and weight teacher supervision on student rollouts.

\section{Method}

\begin{figure}[!t]
\centering
\includegraphics[width=\linewidth]{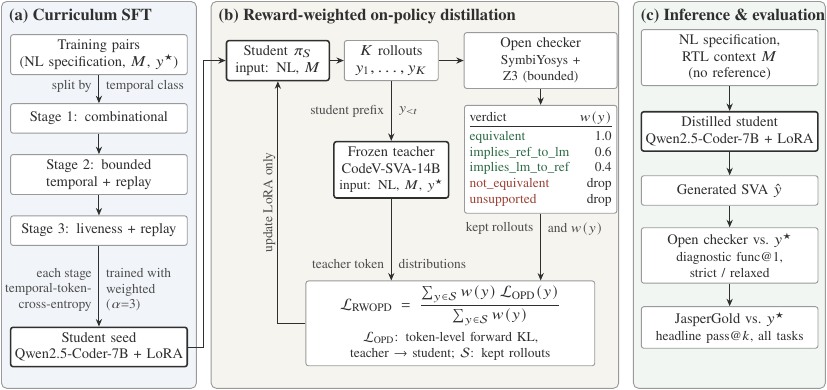}
\caption{RWOPD training and evaluation pipeline.}
\label{fig:overview}
\end{figure}

Given a natural-language specification and RTL context $M$, the student generates an SVA $y$; a reference assertion $y^\star$ is available during training. Figure~\ref{fig:overview} summarizes the two training stages: curriculum SFT initializes the student, then RWOPD distills a frozen teacher on verifier-selected student rollouts. We first define the verifier and its verdicts (\S\ref{sec:pec}), then describe initialization (\S\ref{sec:curriculum}) and the RWOPD objective (\S\ref{sec:opd}). \S\ref{sec:rlvf} defines the separate RLVR control, called reinforcement learning with verifier feedback (RLVF).

\subsection{Open property-equivalence checker}
\label{sec:pec}

\paragraph{Checker inputs and scope.} The open checker compares a generated assertion $y$ with reference $y^\star$ in RTL context $M$. It uses Yosys/SymbiYosys and Z3 \cite{wolf2013yosys,demoura2008z3} for bounded model checking (BMC) \cite{biere1999bmc}. The checker supports a bounded fragment and can abstain. A bounded pass applies to the chosen lowering and checking horizon; it does not prove equivalence for arbitrary SVA.

\paragraph{Verdicts.} The checker tests two directions on a free-input lowering of $M$: (i) assume $y^\star$ and assert $y$, and (ii) assume $y$ and assert $y^\star$. Verdict names distinguish the reference (\emph{ref}) from the language-model candidate (\emph{lm}). A \emph{pass}/\emph{pass} outcome yields \emph{equivalent}. A \emph{pass}/\emph{fail} outcome yields \emph{implies\_ref\_to\_lm}: the reference implies the candidate, so the candidate is weaker (more permissive). A \emph{fail}/\emph{pass} outcome yields \emph{implies\_lm\_to\_ref}: the candidate implies the reference, so it is stronger (more restrictive). A \emph{fail}/\emph{fail} outcome yields \emph{not\_equivalent}. These relations refer to the checked fragment and horizon. An \emph{unsupported} verdict covers unsupported lowering, liveness operators such as \texttt{s\_eventually} and \texttt{s\_until}, and inconclusive runs such as timeouts. Appendix~\ref{app:pecsmoke} gives the verdict matrix and lowering details.

\paragraph{Equivalence, partial relations, and abstention.} Only an \emph{equivalent} verdict establishes agreement in both directions. A one-sided implication establishes only a partial semantic relation. We retain such candidates as lower-weight distillation contexts; we drop unsupported candidates because their relation to the reference is unknown.

\subsection{Curriculum SFT with temporal-token-weighted cross-entropy}
\label{sec:curriculum}

The curriculum SFT stage provides the seed distribution for RWOPD. Two training choices can contribute to the imbalance in \S\ref{sec:gap}: easier temporal classes dominate the data, and uniform cross-entropy gives temporal operators the same weight as other tokens. We address these biases separately with a stratified curriculum over temporal classes \cite{bengio2009curriculum} and an operator-token-weighted cross-entropy.

The curriculum has three stages, one for each temporal class (combinational, bounded temporal, liveness). Later stages draw a fraction of their batches from a uniform replay buffer over earlier-stage pools, to limit forgetting while the model learns harder temporal classes. Per-stage epoch counts, learning-rate decay, validation gates, and replay settings are in Appendix~\ref{app:ttce}.

Temporal-token-weighted cross-entropy (TT-CE) reweights label tokens whose decoded string is a canonical temporal operator (Appendix~\ref{app:ttce}):
\begin{equation}
\mathcal{L}_{\text{TT-CE}} = \frac{1}{T} \sum_{t=1}^{T} w_t \cdot \mathrm{CE}(y_t, \hat{y}_t),
\qquad
w_t = \begin{cases} \alpha & \text{if } y_t \text{ is a temporal-operator token}, \\ 1 & \text{otherwise}, \end{cases}
\end{equation}
where $\mathrm{CE}$ denotes cross-entropy and $\alpha = 3$. The loss reduces to standard cross-entropy at $\alpha = 1$.

\subsection{Reward-weighted OPD}
\label{sec:opd}

RWOPD uses verifier verdicts to select and weight on-policy distillation \cite{agarwal2024onpolicy}. Distillation trains the student to match a teacher's output distribution \cite{hinton2015distilling}. As shown in Figure~\ref{fig:overview}, the frozen teacher receives the reference assertion $y^\star$ in its prompt. At each response position, both models condition on the same student-generated prefix $y_{<t}$; the student prompt contains no reference assertion. Both receive the natural-language prompt and RTL context $M$. The base distillation loss is the forward Kullback--Leibler divergence:
\begin{equation}
\label{eq:opd}
\mathcal{L}_{\text{OPD}}(y) = \frac{1}{|y|}\sum_{t=1}^{|y|}
D_{\mathrm{KL}}\!\left(p_T(\cdot \mid M,y^\star,y_{<t})\,\Vert\,p_S(\cdot \mid M,y_{<t})\right),
\end{equation}
where $p_T$ and $p_S$ are the teacher and student response-token distributions on the shared vocabulary support $[V_{\min}]$; the common natural-language prompt is omitted from the conditioning notation. We compute the loss over response-token positions, with gradients flowing only through the student.

Student rollouts can express incorrect properties. The checker evaluates each completed rollout before distillation, so its semantic relation to the reference determines whether and how strongly its loss contributes to the update.

\paragraph{Partial relations and training coverage.} Exact-equivalence filtering can leave too few rollouts for an update, especially early in training. Accepting one-sided relations increases the available contexts, but treats incomplete semantic matches as weaker evidence than equivalence. We therefore combine a hard gate with relative rollout weights: equivalent candidates receive the largest weight, one-sided candidates receive partial weight, and rejected or unsupported candidates receive zero. This design balances semantic selectivity with training coverage.

\paragraph{Verdict-to-weight mapping.} RWOPD restricts and reward-weights distillation by verifier verdict. For each prompt $p$, the student samples $K$ rollouts at temperature $T$. The open checker (\S\ref{sec:pec}) returns a verdict $\operatorname{verify}(y_i, y^\star, M)$ for each rollout against the canonical reference $y^\star$ in RTL context $M$. The OPD KL is computed only on rollouts whose verdict is \emph{equivalent} or an implication verdict; rollouts returned as \emph{not\_equivalent}, \emph{unsupported}, or syntactically broken are dropped for that step. The kept rollouts are then weighted by the verifier-equivalence reward
\begin{equation}
\label{eq:rwopd_reward}
w(y_i) =
\begin{cases}
1.0 & \text{if }\operatorname{verify}(y_i, y^\star, M) = \mathit{equiv}, \\
0.6 & \text{if }\operatorname{verify}(y_i, y^\star, M) = \mathit{implies}_{\mathit{ref}\to\mathit{lm}}, \\
0.4 & \text{if }\operatorname{verify}(y_i, y^\star, M) = \mathit{implies}_{\mathit{lm}\to\mathit{ref}}, \\
0.0 & \text{otherwise.}
\end{cases}
\end{equation}
We use fixed heuristic weights $1.0/0.6/0.4$; these values are not probabilities of correctness. The $0.6$ weight favors candidates entailed by the reference, while $0.4$ gives less weight to candidates that impose additional constraints. Either one-sided relation can still violate the intended specification.

\paragraph{Normalized distillation objective.} Letting $\mathcal{S}(p) = \{y_i \sim \pi_S(\cdot \mid p) : w(y_i) > 0\}$ be the verifier-passing subset of the $K$ rollouts,
\begin{equation}
\label{eq:rwopd}
\mathcal{L}_{\text{RWOPD}}(p) \;=\;
\frac{\sum_{y \in \mathcal{S}(p)} w(y) \mathcal{L}_{\text{OPD}}(y)}
{\sum_{y \in \mathcal{S}(p)} w(y)}.
\end{equation}
Normalization makes $w(y)$ a relative weight among retained rollouts for the same prompt. For example, an equivalent rollout and a $0.4$-weighted one-sided rollout contribute in a $1:0.4$ ratio; if only one rollout survives, its positive weight cancels. Thus partial weights do not reduce the total update for a prompt containing only one accepted rollout. If $|\mathcal{S}(p)| = 0$, the prompt contributes no gradient in that step. The checker selects rollouts and sets their relative weights; the teacher supplies the token-level targets. We optimize the weighted KL loss without a policy-gradient reward term. The detailed algorithm is in Appendix~\ref{app:opd}.

For vocabulary alignment, we truncate both distributions to the $V_{\min} = 151{,}643$ token identifiers shared by Qwen2.5-Coder-7B and the Qwen3-14B-derived CodeV-SVA-14B. Gradients flow only through the student's LoRA adapter, initialized by the curriculum SFT of \S\ref{sec:curriculum}.

\subsection{Separate RLVR control with verifier rewards}
\label{sec:rlvf}

The RLVF control optimizes the policy using a scalar reward from the same property-equivalence checker. The reward follows the equivalence ordering in Eq.~\ref{eq:rwopd_reward}. Initialized from the curriculum-SFT policy, GRPO \cite{shao2024deepseekmath} consumes
\begin{equation}
\label{eq:reward}
R(\hat{p}, p^\star, M) =
\begin{cases}
1.0  & \text{if }\operatorname{verify}(\hat{p}, p^\star, M) = \text{\emph{equivalent}} \\
0.6  & \text{if \emph{implies\_ref\_to\_lm}} \\
0.4  & \text{if \emph{implies\_lm\_to\_ref}} \\
0.15 & \text{if \emph{unsupported} but syntax\_ok} \\
0.0  & \text{otherwise.}
\end{cases}
\end{equation}
The implication verdicts receive partial credit so GRPO sees an ordered signal for SVAs that are one-sided but not equivalent. The 0.15 floor is only for syntax-valid unsupported outputs; it prevents liveness abstentions from being treated the same as malformed assertions while keeping them far below verified equivalence.

The \texttt{syntax\_ok} branch is defined by the same compile gate used during pool preparation. The gate strips backtick macros, flattens hierarchical paths, normalizes \texttt{disable iff} and ranged-delay forms, and synthesizes a clock-aware free-input wrapper around undeclared signals. The gate determines whether a candidate is syntactically eligible for property-equivalence checking. The full rule list is in Appendix~\ref{app:datapool}.

\FloatBarrier
\section{Experiments}
\label{sec:exp}

We evaluate three questions: how RWOPD compares with existing NL-to-SVA models; how alternative post-training signals perform from a common SFT seed; and how verifier coverage and filtering affect training. The following sections report the protocol, benchmark results, control experiments, and verifier diagnostics.

\subsection{Setup}

\paragraph{Models.} The student is \texttt{Qwen2.5-Coder-7B-Instruct} \cite{hui2024qwen25coder}; the teacher and strongest specialized baseline is \texttt{CodeV-SVA-14B} \cite{wu2026qimeng}, a Qwen3-14B \cite{yang2025qwen3} fine-tune. Both run in bfloat16 (bf16) precision on a single H200, with teacher forward passes accounting for most OPD runtime.

\paragraph{Training corpus and benchmark separation.} We train the main SFT, OPD, and RWOPD models on the CodeV-SVA-14B training split \cite{wu2026qimeng} and evaluate on NL2SVA-Human and NL2SVA-Machine. We retain the released CodeV-SVA/FVEval separation between training and evaluation rows. Appendix~\ref{app:datapool} describes auxiliary pools used for compile-gate and verifier-reward diagnostics. The appendix experiments using NL2SVA-Machine-derived paraphrases are negative controls; those pools are excluded from the main RWOPD training data.

\paragraph{Evaluation.} We evaluate on the NL2SVA-Human and NL2SVA-Machine splits under the FVEval protocol, generating $n{=}32$ candidates per task under the same 32K-token generation budget. JasperGold provides the headline correctness metric; the open checker provides the bounded-fragment training signal and auxiliary diagnostics (\S\ref{sec:pec}). The same compile gate used for training-pool preparation is also ablated on the 39{,}914-row master pool: raw lint with the Synopsys Verilog compiled simulator (VCS) passes 26.5\% of candidates, while normalization plus the clock-aware wrapper passes 78.6\% (Appendix~\ref{app:datapool}).

\paragraph{Metrics.} For each task $i$, we estimate \emph{pass@$k$} from $n{=}32$ samples using $\mathrm{pass@}k_i = 1 - \binom{n - c_i}{k} / \binom{n}{k}$, where $c_i$ is the number of correct samples, and average over tasks. Table~\ref{tab:main} uses Cadence JasperGold's \texttt{prop\_eq\_checker} for consistency with prior work. \emph{compile@1} is the fraction of tasks whose first sample passes the clock-aware free-input VCS \texttt{-sverilog -assert svaext} compile gate. The open-checker metrics report functional accuracy from one generated candidate per task (func@1). \emph{Strict checker func@1} counts only \emph{equivalent} verdicts; \emph{relaxed checker func@1} also counts either one-sided implication. The relaxed checker score measures an accepted relation, not equivalence accuracy. Human checker scores use the fixed 67-task evaluable subset. These evaluation criteria are independent of the training filter.

\paragraph{Comparison methods.} RS-SFT, GRPO, identity preference optimization (IPO), unfiltered OPD, and RWOPD are alternative branches from the same curriculum-SFT initialization. RS-SFT trains on accepted student responses with cross-entropy and no teacher; GRPO optimizes verifier rewards; IPO learns from preference pairs labeled by the checker; OPD distills the teacher on unfiltered student rollouts; RWOPD adds verdict-based selection and weighting. The broader GRPO/IPO campaign also varies reward shape, data pools, and hyperparameters (Appendix~\ref{app:grpo}); these runs test several training choices and do not isolate a single RWOPD component.

\subsection{Main results}

\begin{table}[t]
\centering
\caption{Main results (percent; higher is better). JasperGold pass@$k$ uses $n{=}32$ candidates per task. RS-SFT, GRPO, OPD, and RWOPD each start from the curriculum-SFT seed; rows are alternative branches, not successive training stages.}
\label{tab:main}
\setlength{\tabcolsep}{3.5pt}
\resizebox{\linewidth}{!}{%
\begin{tabular}{lcccccccc}
\toprule
 & \multicolumn{4}{c}{NL2SVA-Human} & \multicolumn{4}{c}{NL2SVA-Machine} \\
\cmidrule(lr){2-5}\cmidrule(lr){6-9}
Model & compile@1 & pass@1 & pass@5 & pass@10 & compile@1 & pass@1 & pass@5 & pass@10 \\
\midrule
\multicolumn{9}{l}{\textit{Advanced general-purpose models}} \\
DeepSeek-R1-671B           & 78.5 & 74.6 & 90.3 & 90.4 & 84.3 & 81.0 & 93.3 & 94.3 \\
GPT-5                      & 75.9 & 71.8 & 90.2 & 92.7 & 85.3 & 81.8 & 93.2 & 94.3 \\
DeepSeek-V3.1-671B         & 67.1 & 63.1 & 81.4 & 84.9 & 87.0 & 83.8 & 92.9 & 93.6 \\
GPT-4o                     & 68.4 & 64.1 & 75.2 & 78.1 & 72.0 & 68.5 & 81.3 & 83.7 \\
\midrule
\multicolumn{9}{l}{\textit{Specialized RTL generation models}} \\
RTLCoder-DS-v1.1-6.7B      & 30.4 & 25.9 & 58.8 & 65.8 & 25.0 & 21.7 & 54.8 & 60.8 \\
CodeV-R1-Qwen-7B           & 29.1 & 25.2 & 55.8 & 61.6 & 41.0 & 37.4 & 76.6 & 83.0 \\
CodeV-SVA-8B               & 75.9 & 72.0 & 88.8 & 90.4 & 87.0 & 83.5 & 96.3 & 97.2 \\
\textbf{CodeV-SVA-14B (teacher)} & \textbf{79.7} & \textbf{75.8} & \textbf{89.4} & \textbf{90.4} & \textbf{87.3} & \textbf{84.0} & \textbf{94.9} & \textbf{95.8} \\
\midrule
\multicolumn{9}{l}{\textit{Open-source foundation models}} \\
Qwen3-8B                   & 36.7 & 32.3 & 71.6 & 74.0 & 50.0 & 46.1 & 88.0 & 90.5 \\
Qwen3-14B                  & 65.8 & 61.6 & 86.1 & 87.7 & 79.0 & 75.3 & 92.7 & 94.3 \\
\midrule
\multicolumn{9}{l}{\textit{Ours (7B+LoRA student, 0.53\% trainable)}} \\
Qwen2.5-Coder-7B zero-shot                       & 29.1 & 25.3 & 58.2 & 65.8 & 25.0 & 21.7 & 54.7 & 60.7 \\
Curriculum SFT seed, replay 50\%              & 74.7 & 70.9 & 87.3 & 88.6 & 86.7 & 83.3 & 96.0 & 97.0 \\
SFT seed $\to$ RS-SFT (no teacher)         & 75.9 & 73.4 & 88.0 & 89.2 & 87.0 & 83.7 & 96.0 & 97.0 \\
SFT seed $\to$ GRPO (best GRPO config)         & 78.5 & 74.7 & 88.6 & 89.9 & 87.0 & 83.7 & 94.7 & 95.7 \\
SFT seed $\to$ OPD (14B teacher)              & 84.8 & 77.2 & 89.9 & 91.1 & 89.3 & 85.3 & 96.3 & 97.3 \\
\rowcolor{blue!8}
\textbf{SFT seed $\to$ RWOPD (14B teacher)}  & \textbf{91.1} & \textbf{78.8} & \textbf{92.4} & \textbf{93.4} & \textbf{90.7} & \textbf{87.0} & \textbf{97.9} & \textbf{98.8} \\
\bottomrule
\end{tabular}%
}
\end{table}

\begin{table}[t]
\centering
\caption{Task-level bootstrap 95\% confidence intervals for pass@$k$.}
\label{tab:bootstrap_ci}
\small
\resizebox{\linewidth}{!}{%
\begin{tabular}{lcccccc}
\toprule
Model & Human pass@1 & Human pass@5 & Human pass@10 & Machine pass@1 & Machine pass@5 & Machine pass@10 \\
\midrule
CodeV-SVA-14B  & 67.1--84.8 & 82.3--96.2 & 83.5--97.5 & 80.0--88.3 & 92.3--97.3 & 93.7--98.0 \\
SFT-50\% seed  & 60.8--81.0 & 79.7--94.9 & 82.3--96.2 & 79.0--87.7 & 93.7--98.3 & 95.0--99.0 \\
OPD            & 68.4--86.1 & 83.5--96.2 & 84.8--97.5 & 81.3--89.3 & 94.3--98.3 & 95.7--99.0 \\
\textbf{RWOPD} & \textbf{69.6--87.3} & \textbf{86.1--98.7} & \textbf{87.3--98.7} & \textbf{83.3--90.7} & \textbf{96.3--99.7} & \textbf{97.7--100.0} \\
\bottomrule
\end{tabular}
}
\end{table}

\paragraph{Overall performance.} RWOPD has the highest point estimate in every column of Table~\ref{tab:main}. Human/Machine pass@1 is 78.8/87.0\%, compared with 75.8/84.0\% for the CodeV-SVA-14B teacher and 77.2/85.3\% for unfiltered OPD. The absolute pass@1 gains over the teacher are 3.0\% on each benchmark, and those over OPD are 1.6/1.7\%. compile@1 is 91.1/90.7\%. These gains use a 7B student with 0.53\% trainable parameters.

\paragraph{Comparison of post-training methods.} Curriculum SFT raises Human/Machine pass@1 from 25.3/21.7\% to 70.9/83.3\%. Starting independently from that seed, RS-SFT reaches 73.4/83.7\% and GRPO reaches 74.7/83.7\%. OPD and RWOPD reach higher point estimates using teacher distributions. These comparisons support the usefulness of dense teacher supervision in this setting; they do not estimate additive contributions of RS-SFT, GRPO, OPD, and RWOPD to one training sequence.

\paragraph{Confidence intervals.}
\label{sec:bootstrap-ci}

Table~\ref{tab:bootstrap_ci} reports task-level bootstrap 95\% confidence intervals for pass@$k$. We resample evaluation tasks with replacement, recompute $\mathrm{pass@}k_i = 1 - \binom{n-c_i}{k}/\binom{n}{k}$ for each task, and average the estimates over tasks in each of 10{,}000 replicates. Marginal intervals overlap between adjacent methods. RWOPD has the highest point estimate in every column on both benchmarks.

\paragraph{Performance by temporal class.} Figure~\ref{fig:per_class_machine} compares the 14B teacher with the 7B RWOPD student by reference class on Machine pass@1. Both models score 92.6\% on combinational assertions, the majority class. RWOPD yields absolute gains of 7.5\% on bounded temporal assertions and 11.4\% on liveness assertions. JasperGold evaluates all three classes, including liveness, for which the open checker abstains during training. We omit the Human class breakdown because its small bounded temporal and liveness subsets make the rates sensitive to a single task.

\begin{figure}[t]
\centering
\begin{minipage}[t]{0.48\linewidth}
\centering
\begin{tikzpicture}
\begin{axis}[
    width=\linewidth,
    height=4.6cm,
    ybar,
    bar width=4pt,
    enlarge x limits=0.18,
    symbolic x coords={Per-RTL, +AST, Equivalence, Hybrid},
    xticklabels={Per-RTL,+AST,{property\\equivalence},Hybrid},
    xtick=data,
    xticklabel style={font=\tiny, align=center, rotate=20, anchor=north east},
    ymin=0, ymax=1.18,
    ylabel={Reward value},
    legend style={at={(0.5,-0.35)}, anchor=north, legend columns=4, font=\scriptsize, draw=gray!50},
    legend cell align=left,
    tick label style={font=\footnotesize},
    label style={font=\footnotesize},
]
\addplot[fill=green!50!black, draw=green!40!black] coordinates {(Per-RTL, 0.510) (+AST, 1.000) (Equivalence, 1.000) (Hybrid, 1.000)};
\addlegendentry{golden}
\addplot[fill=gray!50, draw=black!50] coordinates {(Per-RTL, 0.200) (+AST, 0.000) (Equivalence, 0.000) (Hybrid, 0.000)};
\addlegendentry{vacuous}
\addplot[fill=red!60!black, draw=red!50!black] coordinates {(Per-RTL, 0.510) (+AST, 0.867) (Equivalence, 0.000) (Hybrid, 0.433)};
\addlegendentry{swap}
\addplot[fill=orange!75!black, draw=orange!60!black] coordinates {(Per-RTL, 0.460) (+AST, 0.933) (Equivalence, 0.400) (Hybrid, 0.667)};
\addlegendentry{flip}
\end{axis}
\end{tikzpicture}
\captionof{figure}{Reference and mutation rewards. AST denotes abstract syntax tree.}
\label{fig:rewardstudy}
\end{minipage}
\hfill
\begin{minipage}[t]{0.48\linewidth}
\centering
\begin{tikzpicture}
\begin{axis}[
    width=\linewidth,
    height=4.6cm,
    ybar,
    bar width=7pt,
    enlarge x limits=0.4,
    xtick=data,
    symbolic x coords={combinational, bounded temporal, liveness},
    xticklabels={combinational,{bounded\\temporal},liveness},
    xticklabel style={font=\scriptsize, align=center},
    ymin=0, ymax=130,
    ylabel={pass@1 (\%)},
    nodes near coords,
    nodes near coords style={font=\tiny, rotate=90, anchor=west, /pgf/number format/.cd, fixed, fixed zerofill, precision=1},
    legend style={at={(0.5,-0.35)}, anchor=north, legend columns=2, font=\scriptsize, draw=gray!50},
    tick label style={font=\footnotesize},
    label style={font=\footnotesize},
]
\addplot[fill=gray!55, draw=black!60] coordinates {(combinational, 92.6) (bounded temporal, 75.5) (liveness, 35.5)};
\addlegendentry{14B teacher}
\addplot[fill=blue!50!black!60, draw=blue!60!black] coordinates {(combinational, 92.6) (bounded temporal, 83.0) (liveness, 46.9)};
\addlegendentry{RWOPD (ours)}
\end{axis}
\end{tikzpicture}
\captionof{figure}{Machine pass@1 by temporal class.}
\label{fig:per_class_machine}
\end{minipage}
\end{figure}

\subsection{Ablation study}
\label{sec:exp-analysis}

\subsubsection{Effect of teacher supervision}
\label{sec:opd-vs-rlvf}

The common-seed comparison asks whether verifier feedback is more useful for selecting teacher supervision or as the sole post-training feedback. Figure~\ref{fig:opd_vs_rlvf} (left) compares the SFT seed and five alternative branches. RS-SFT raises Human JasperGold pass@1 from 70.9\% to 73.4\%, but leaves strict/relaxed checker scores at 29.9/55.2\%. RWOPD reaches 78.8\% pass@1 and strict/relaxed checker scores of 38.8/65.7\%. This supports combining teacher supervision with verifier selection, although RS-SFT versus RWOPD changes both the target and the loss and is not a pure filter ablation.

Figure~\ref{fig:rewardstudy} compares mean rewards on 50 reference SVAs and three mutation types: vacuous assertions, swapped implication operands, and changed Boolean conditions. The reward designs are per-RTL verification, per-RTL verification with abstract syntax tree (AST) similarity, reference-grounded property-equivalence checking, and a hybrid reward. Under free inputs, per-RTL verification gives the reference and swap mutation the same mean reward, leaving GRPO advantage normalization little signal. Reference-grounded property-equivalence checking increases the mean reward gap between reference and swapped assertions to 1.00. This reward distinguishes the tested semantic mutations, but the RLVF runs do not improve strict held-out checker scores over the SFT seed.

The eight GRPO configurations and two IPO configurations test reward shapes, training pools, and optimization settings (Appendix~\ref{app:grpo}). Figure~\ref{fig:opd_vs_rlvf} reports strict/relaxed Human checker scores of 29.9/55.2\% for the GRPO comparison and 28.4/53.7\% for IPO. None exceeds the seed's 29.9\% strict Human checker score, even though the best GRPO branch improves JasperGold pass@1 to 74.7\%. These metric-dependent outcomes limit the conclusion to the tested verifier, data, and 7B+LoRA regime. The matched OPD-versus-RWOPD comparison evaluates filtering and weighting together; it does not isolate the numerical weight values.

\subsubsection{Effect of filtering policy}
\label{sec:filter-acceptance}

The fraction of rollouts accepted as \emph{equivalent} or one-sided implication increases from roughly 0.32 at the SFT-50\% seed to about 0.80 by step 200. More rollouts therefore contribute to the distillation update as training proceeds. We compare three filter policies with the same seed, teacher, LoRA rank, and budget (Figure~\ref{fig:opd_vs_rlvf}, middle and right). Unfiltered OPD applies teacher KL to every syntactically valid rollout. Equivalence-only OPD retains only \emph{equivalent} rollouts. RWOPD retains equivalent and one-sided rollouts with the weights in Eq.~\ref{eq:rwopd_reward}; the default is $K{=}8$.

Equivalence-only OPD has fewer accepted rollouts at small $K$: Human pass@1 is 71.5 at $K{=}1$ and recovers to 78.5 at $K{=}8$. Unfiltered OPD stays near 78 pass@1 across $K$. RWOPD gives the highest reported scores in this comparison. Training filters and evaluation criteria are independent. \emph{Equivalence-only OPD} trains only on equivalent rollouts, but we evaluate its outputs using both the strict criterion (equivalence only) and the relaxed criterion (equivalence or either implication). The relaxed score therefore applies even when training used the equivalence-only filter. The middle and right panels of Figure~\ref{fig:opd_vs_rlvf} plot these held-out scores against optimizer step on Human and Machine, respectively. Color and marker identify the training method; solid and dashed lines show strict and relaxed checker scores, respectively. All Human checker scores use the fixed 67-task subset. This comparison does not isolate the effect of the $0.6/0.4$ partial-weight asymmetry; a comparison with symmetric partial weights would be needed to assess the choice of constants.

\begin{figure}[t]
\centering
\begin{tikzpicture}
\begin{groupplot}[
    group style={group size=3 by 1, horizontal sep=1.1cm},
    width=0.29\linewidth,
    height=3.8cm,
    grid=both,
    grid style={line width=.1pt, draw=gray!30},
    tick label style={font=\footnotesize},
    label style={font=\footnotesize},
    title style={font=\footnotesize},
    legend cell align=left,
    legend style={font=\scriptsize, draw=gray!50, inner sep=2pt, row sep=1pt,
        nodes={text height=1.5ex, text depth=0.4ex, inner ysep=0.5pt}},
]
\nextgroupplot[
    title={Common-seed branches},
    xtick={1,2,3,4,5,6},
    xticklabels={SFT, RS-SFT, GRPO, IPO, OPD, RWOPD},
    xticklabel style={font=\tiny, rotate=30, anchor=north east},
    xmin=0.6, xmax=6.4,
    ymin=20, ymax=92,
    ylabel={func@1 / pass@1 (\%)},
    legend style={at={(0.5,0)}, yshift=-33pt, anchor=north, legend columns=2},
]
\addplot[color=blue!60!black, mark=*, only marks, mark size=2pt]
  coordinates {(1, 29.9) (2, 29.9) (3, 29.9) (4, 28.4) (5, 31.3) (6, 38.8)};
\addlegendentry{Human strict}
\addplot[color=blue!60!black, mark=square*, only marks, mark size=2pt]
  coordinates {(1, 55.2) (2, 55.2) (3, 55.2) (4, 53.7) (5, 62.7) (6, 65.7)};
\addlegendentry{Human relaxed}
\addplot[color=red!70!black, mark=triangle*, only marks, mark size=2pt]
  coordinates {(1, 70.9) (2, 73.4) (3, 74.7) (4, 72.2) (5, 77.2) (6, 78.8)};
\addlegendentry{Human pass@1}
\addplot[color=red!70!black, mark=diamond*, only marks, mark size=2pt]
  coordinates {(1, 83.3) (2, 83.7) (3, 83.7) (4, 83.3) (5, 85.3) (6, 87.0)};
\addlegendentry{Machine pass@1}

\nextgroupplot[
    title={Trajectory (Human)},
    xlabel={Training step},
    ylabel={Checker func@1 (\%)},
    xmin=0, xmax=210,
    ymin=15, ymax=72,
    legend style={at={(1.4,0)}, yshift=-33pt, anchor=north, legend columns=3,
        /tikz/every even column/.append style={column sep=4pt}},
]
\addplot[forget plot, color=red!70!black, mark=square*, thick]
  coordinates {(50,31.3) (100,22.4) (150,25.4) (200,22.4)};

\addplot[forget plot, color=red!70!black, mark=square, thick, dashed]
  coordinates {(50,62.7) (100,59.7) (150,59.7) (200,59.7)};

\addplot[forget plot, color=orange!80!black, mark=triangle*, thick]
  coordinates {(10,28.4) (25,29.9) (50,32.8) (100,34.3) (150,35.8) (200,34.3)};

\addplot[forget plot, color=orange!80!black, mark=triangle, thick, dashed]
  coordinates {(10,50.7) (25,53.7) (50,56.7) (100,61.2) (150,62.7) (200,61.2)};

\addplot[forget plot, color=blue!60!black, mark=*, thick]
  coordinates {(10,29.9) (25,32.8) (50,34.3) (100,37.3) (150,38.8) (200,37.3)};

\addplot[forget plot, color=blue!60!black, mark=o, thick, dashed]
  coordinates {(10,53.7) (25,56.7) (50,59.7) (100,62.7) (150,65.7) (200,64.2)};

\addlegendimage{red!70!black,mark=square*,only marks}
\addlegendentry{Unfiltered OPD}
\addlegendimage{orange!80!black,mark=triangle*,only marks}
\addlegendentry{Equiv.-only OPD}
\addlegendimage{blue!60!black,mark=*,only marks}
\addlegendentry{RWOPD}
\addlegendimage{black,thick}
\addlegendentry{Eval: strict}
\addlegendimage{black,thick,dashed}
\addlegendentry{Eval: relaxed}

\nextgroupplot[
    title={Trajectory (Machine)},
    xlabel={Training step},
    ylabel={Checker func@1 (\%)},
    xmin=0, xmax=210,
    ymin=30, ymax=60,
]
\addplot[color=red!70!black, mark=square*, thick]
  coordinates {(50,38.7) (100,38.7) (150,38.3) (200,34.0)};
\addplot[color=red!70!black, mark=square, thick, dashed]
  coordinates {(50,48.0) (100,48.0) (150,48.0) (200,41.3)};
\addplot[color=orange!80!black, mark=triangle*, thick]
  coordinates {(10,36.0) (25,38.0) (50,39.7) (100,41.0) (150,42.0) (200,41.0)};
\addplot[color=orange!80!black, mark=triangle, thick, dashed]
  coordinates {(10,47.0) (25,49.0) (50,50.7) (100,52.0) (150,53.0) (200,52.0)};
\addplot[color=blue!60!black, mark=*, thick]
  coordinates {(10,38.0) (25,40.0) (50,41.0) (100,42.7) (150,44.0) (200,43.0)};
\addplot[color=blue!60!black, mark=o, thick, dashed]
  coordinates {(10,49.0) (25,51.0) (50,52.7) (100,54.0) (150,55.0) (200,54.0)};
\end{groupplot}
\end{tikzpicture}
\caption{Post-training controls and open-checker learning curves. Strict and relaxed scores use the open property-equivalence checker.}
\label{fig:opd_vs_rlvf}
\end{figure}

\subsubsection{Verifier consistency}
\label{sec:errorbreakdown}

Reference-vs-reference checking abstains on 12 NL2SVA-Human tasks (11 liveness cases and one timeout on a combinational task), leaving a fixed 67-task subset for the Human diagnostics from the open checker. JasperGold evaluates the full benchmark. We compare strict verdicts from the open checker with JasperGold on greedy outputs from the SFT-50\% seed. One-sided implications are excluded from this equivalence audit. On NL2SVA-Human, JasperGold confirms all 20 strict accepts from the open checker. We extend the audit to all 300 NL2SVA-Machine outputs: JasperGold confirms all 124 strict accepts there as well. The checker abstains, reports one-sided implication, or rejects the remaining 176 Machine outputs. Across the 379 outputs from both benchmarks, JasperGold disagrees with none of the 144 strict accepts from the open checker. It also accepts outputs that the checker labels one-sided, unsupported, or rejected. This difference supports using JasperGold for the main pass@$k$ results and the open checker as a bounded training-time check; the audit covers one checkpoint.

\subsubsection{Effect of training duration}
\label{sec:budget}

We select checkpoints by the open checker's strict score on a validation slice of the CodeV-SVA training split. This slice is disjoint from the gradient-training prompts and both NL2SVA benchmarks. The protocol selects step 150. Figure~\ref{fig:opd_vs_rlvf} shows post-hoc Human and Machine score trajectories from the open checker over 200 steps for unfiltered OPD, equivalence-only OPD, and RWOPD. These benchmark curves were not used for selection. Both filtered variants improve through step 150 and then decline slightly by step 200. RWOPD peaks at strict/relaxed scores of 38.8/65.7 on Human and 44.0/55.0 on Machine; equivalence-only OPD scores slightly lower. Unfiltered OPD reaches its best plotted scores at step 50, then declines or plateaus. These trajectories motivate validation-based selection but do not establish stability across training seeds.


\section{Conclusion}
\label{sec:limitations}

We study temporal structure and semantic correctness in NL-to-SVA generation, and use a formal verifier to select and weight dense teacher supervision. The resulting curriculum-SFT $\to$ RWOPD pipeline has the highest reported point estimates across the two benchmarks in Table~\ref{tab:main}. Independent RS-SFT and GRPO controls improve JasperGold pass@1 over the SFT seed but remain below RWOPD; their strict Human checker scores do not exceed the seed. These results favor verifier-guided distillation within the tested 7B+LoRA setting. The exact partial-weight values remain heuristic, liveness gains cannot be attributed to direct supervision from the open checker, and broader comparisons with recent joint or sequential OPD--RLVR methods remain open.

\begin{ack}
Funding. This research/project is supported by the National Research Foundation, Singapore and Infocomm Media Development Authority under its Trust Tech Funding Initiative. Any opinions, findings and conclusions or recommendations expressed in this material are those of the author(s) and do not reflect the views of National Research Foundation, Singapore and Infocomm Media Development Authority. Competing interests. The authors declare no competing interests.
\end{ack}


{\small
\bibliographystyle{plain}
\bibliography{reference}
}


\newpage
\appendix

\section{Terminology and benchmark composition}
\label{app:terminology}

\paragraph{Terminology.} Table~\ref{tab:glossary} summarizes the main terms used in the paper. We use the combinational, bounded temporal, and liveness classes only as a diagnostic description of temporal-operator structure; semantic correctness is measured by JasperGold pass@$k$ and, for the open bounded fragment, by verdicts from the open checker.

\begin{table}[h]
\centering
\caption{Glossary of recurring terms.}
\label{tab:glossary}
\small
\begin{tabular}{p{0.18\linewidth}p{0.74\linewidth}}
\toprule
Term & Meaning in this paper \\
\midrule
SVA & SystemVerilog assertion: a formal assertion over RTL behavior, used to specify safety, bounded-time, or liveness properties. \\
OPD & on-policy distillation: the student samples its own rollouts and learns from a frozen teacher's token distributions. \\
RWOPD & reward-weighted OPD: distillation from a reference-conditioned teacher to the student on verifier-selected rollouts, using normalized verdict-based weights. \\
RLVF & reinforcement learning with verifier feedback: a baseline that maps verifier verdicts to sparse rollout-level rewards. \\
Open checker & Property-equivalence checker: the open SymbiYosys+Z3 checker that compares a generated SVA with the reference by bounded property-equivalence checks. \\
Relaxed score & Open-checker score: a diagnostic score that counts strict equivalence and accepted one-sided implication verdicts as passable. \\
Temporal complexity & Temporal complexity level: a diagnostic class assigned from the strongest temporal operator in an SVA; not a semantic correctness metric. \\
Temporal classes & Combinational, bounded temporal, and liveness assertions, ordered by increasing operator complexity. \\
JasperGold & Cadence JasperGold \texttt{prop\_eq\_checker}: the industrial formal oracle used for complete headline pass@$k$ evaluation. \\
Compile gate & normalization and lint gate: SVA/RTL preprocessing that normalizes syntax and checks whether a candidate can enter verifier-based diagnostics. \\
\bottomrule
\end{tabular}
\end{table}

\paragraph{Held-out benchmark composition.} Table~\ref{tab:benchmark_composition} reports the temporal-class composition of the two held-out evaluation files. These counts describe the reference SVA structure and are used to contextualize aggregate pass@$k$; they do not by themselves measure whether a generated assertion is semantically correct.

\begin{table}[h]
\centering
\caption{Held-out benchmark composition by temporal class.}
\label{tab:benchmark_composition}
\small
\begin{tabular}{lrrrr}
\toprule
Held-out file & Rows & combinational & bounded temporal & liveness \\
\midrule
\texttt{nl2sva\_human.jsonl}  & 79  & 62  & 6  & 11 \\
\texttt{nl2sva\_machine.jsonl} & 300 & 189 & 94 & 17 \\
\bottomrule
\end{tabular}
\end{table}

\section{Pipeline and reproducibility}
\label{app:pipeline}

Training and evaluation run in a single conda environment (PyTorch 2.10, Transformers 4.57, Transformers reinforcement learning 0.27, parameter-efficient fine-tuning 0.18) plus OSS CAD Suite (Yosys + yosys-slang + SymbiYosys + Z3), for the property-equivalence checker. Stage runtimes on a single H200 GPU are: NL fill $\sim$1.9\,h; checker self-coverage $\sim$30\,s; checker evaluation per model $\sim$45\,s; curriculum SFT $\sim$1.5\,h.

\section{On-policy distillation implementation}
\label{app:opd}

\paragraph{Objective.} RWOPD minimizes the weighted forward-KL loss on response tokens of accepted student rollouts (Eq.~\ref{eq:rwopd}). Each step proceeds as follows:
(a) sample $K$ rollouts $\{y_1, \ldots, y_K\}$ from the student at $T = 1.0$, $\text{top-}p = 0.95$, max 1024 response tokens, with a left-padded prompt at most 4096 tokens;
(b) verify each rollout via $\operatorname{verify}(y_i, y^\star, M)$ on the free-input lowering of the RTL context, capped at the same $60$\,s BMC budget as evaluation;
(c) assign rollout weights $w_i$ as in Eq.~\ref{eq:rwopd_reward}: 1.0 for \emph{equivalent}, 0.6 for \emph{implies\_ref\_to\_lm}, 0.4 for \emph{implies\_lm\_to\_ref}, and 0 otherwise; form $\mathcal{S}(p) = \{y_i : w_i > 0\}$; if $|\mathcal{S}(p)| = 0$ skip the prompt;
(d) for each $y \in \mathcal{S}(p)$, append the same student rollout to the reference-conditioned teacher prompt and the reference-free student prompt; compute teacher logits without gradients and student logits with gradients, and align the logits at response-token positions using each prompt's own length; both prompts include RTL context $M$; right-truncate both logit tensors to vocabulary index $V_{\min} = 151{,}643$ (the Qwen2.5 / Qwen3 byte-pair encoding intersection);
(e) restrict the loss to response-token positions only and take the normalized reward-weighted average across $\mathcal{S}(p)$;
(f) backpropagate the loss and take an adaptive moment estimation step with decoupled weight decay (AdamW).
We use $K = 8$ unless stated otherwise, matching the default in \S\ref{sec:filter-acceptance}.

\paragraph{Hyperparameters.} LoRA \cite{hu2022lora} rank $r{=}16$, $\alpha{=}32$, dropout 0.05, target modules \texttt{q\_proj, k\_proj, v\_proj, o\_proj, gate\_proj, up\_proj, down\_proj}, no bias. Optimizer: AdamW with $\eta = 5\!\times\!10^{-6}$, weight decay 0.01, cosine schedule with 5\% warmup, and gradient norm clip 1.0. The prompt batch size is 1: each prompt yields $K$ rollouts, and an optimizer step uses the normalized weighted loss over its retained set $\mathcal{S}(p)$ (Eq.~\ref{eq:rwopd}). Prompts with an empty retained set are skipped. Seed adapter: the curriculum-SFT-50\% reasoning-augmented LoRA adapter at step 12{,}000 of a 14{,}975-prompt SFT run (\S\ref{sec:curriculum}).

\paragraph{Vocabulary alignment.} CodeV-SVA-14B is a Qwen3-14B fine-tune; the student is Qwen2.5-Coder-7B-Instruct. The two tokenizers share 151{,}643 token identifiers in identical positions; the Qwen3 vocabulary's longer suffix is unused on this task. We truncate both logit tensors to $V_{\min} = 151{,}643$ before the softmax and have not observed truncation-induced degradation on smoke prompts.

\paragraph{Checkpoint selection.} We select checkpoints by the open checker's strict score on a validation slice of the CodeV-SVA training split. The slice is disjoint from the gradient-training data and both NL2SVA benchmarks.

\section{Temporal class definitions and classifier validation}
\label{app:tcl}

\paragraph{Class definitions.} The three temporal classes of \S\ref{sec:gap} are defined by the maximum-complexity temporal operator in an SVA's abstract syntax tree:

\noindent\textbf{Combinational.} No temporal operator; sampled-value functions only (\texttt{\$rose}, \texttt{\$fell}, \texttt{\$stable}). Decidable on a single sampled cycle.

\noindent\textbf{Bounded temporal.} Fixed delay (\texttt{\#\#N}, \texttt{[*N]}), ranged delay (\texttt{\#\#[a:b]}, \texttt{[*a:b]}), or sequence implication (\texttt{|->}, \texttt{|=>}, \texttt{throughout}, \texttt{within}, \texttt{intersect}). Decidable in bounded model checking.

\noindent\textbf{Liveness.} \texttt{s\_eventually}, \texttt{s\_until}, \texttt{s\_always}, \texttt{until\_with}. Requires liveness verification; bounded model checking does not decide.

\paragraph{Classifier.} The regular-expression-based AST classifier strips comments and label prefixes, scans for the canonical operators, and returns the highest class present. It reaches 100\% accuracy on a 90-SVA hand-labeled corpus (30 per class). Any \texttt{|->} assigns at least the bounded temporal class, even if the rest of the body is combinational. Any \texttt{s\_eventually}, \texttt{s\_until}, \texttt{s\_always}, or \texttt{until\_with} assigns the liveness class regardless of the bounded operators present. The 32 unit tests cover comment stripping, label prefixes, \texttt{\#\#[a:b]} ranges, and sampled-value functions such as \texttt{\$rose}/\texttt{\$fell}, which remain combinational.

\section{Curriculum schedule and temporal-token-weighted cross-entropy implementation}
\label{app:ttce}

\paragraph{Curriculum schedule.} The 3-stage curriculum of \S\ref{sec:curriculum} aligns one stage with one temporal class. Stage 1 trains on combinational assertions for 3 epochs at $2\times 10^{-5}$ to a validation temporal-class accuracy target of $\geq 0.85$ with no replay; stage 2 on bounded temporal assertions for 5 epochs at $1\times 10^{-5}$ to $\geq 0.65$ with 50\% replay; stage 3 on liveness assertions for 6 epochs at $8\times 10^{-6}$ to $\geq 0.50$ with 50\% replay. We use AdamW with weight decay 0.01, bf16 mixed precision, gradient checkpointing, and linear learning-rate decay within each stage. Replay rate affects performance on combinational assertions: with 20\% replay the after-stage-3 model reaches checker-confirmed equivalence on 13 of 62 combinational tasks; with 50\% replay it reaches 18. The open checker abstains on all 11 Human liveness tasks (\S\ref{sec:errorbreakdown}), so this checker comparison measures retention of performance on combinational assertions. We evaluate semantic correctness on liveness assertions with JasperGold, as in Figure~\ref{fig:per_class_machine}.

\paragraph{Operator vocabulary.} The reweighted-token set $\mathcal{O}_{\text{temp}}$ is the canonical temporal operator vocabulary:
\begin{multline*}
\mathcal{O}_{\text{temp}} = \{\,\texttt{\#\#},\ \texttt{[*},\ \texttt{[=},\ \texttt{|->},\ \texttt{|=>},\ \texttt{until},\ \texttt{eventually},\ \texttt{s\_eventually}, \\
\texttt{s\_until},\ \texttt{s\_always},\ \texttt{throughout},\ \texttt{within},\ \texttt{intersect},\ \texttt{\$rose},\ \texttt{\$fell}\,\}.
\end{multline*}
The membership test runs on the decoded string of each label token, which handles byte-pair encoding merges that split \texttt{|->} into multiple sub-tokens (any sub-token whose decode contains an element of $\mathcal{O}_{\text{temp}}$ is reweighted).

\paragraph{Reference implementation.} The reweighting in \S\ref{sec:curriculum} is implemented per-token at the loss layer:
\begin{lstlisting}[language=Python]
TEMPORAL_OPS = {"##","[*","[=","|->","|=>","until","eventually",
                "s_eventually","s_until","s_always","throughout",
                "within","intersect","$rose","$fell"}

def temporal_weighted_ce(logits, labels, tok, alpha=3.0):
    base = F.cross_entropy(logits, labels, reduction="none")
    w = torch.ones_like(base)
    for i, tid in enumerate(labels):
        if any(op in tok.decode([tid]) for op in TEMPORAL_OPS):
            w[i] = alpha
    return (base * w).mean()
\end{lstlisting}
A 7-test NumPy unit suite verifies (a) reduction to standard CE at $\alpha=1$, (b) correct reweighting on \texttt{\#\#3}, \texttt{|->}, \texttt{s\_eventually}, and (c) no double-counting when a token decodes to two operators.

\section{Property-equivalence checker implementation and validation}
\label{app:pecsmoke}

\paragraph{Verdict matrix.} Each checker call runs two BMC instances on the free-input RTL: \textcircled{1} $\texttt{assume}(y^\star)\wedge\texttt{assert}(y)$ and \textcircled{2} $\texttt{assume}(y)\wedge\texttt{assert}(y^\star)$. We map the \emph{pass}/\emph{fail} pair to a verdict as in Table~\ref{tab:pec_verdict}.

\begin{table}[h]
\centering
\caption{Checker verdict matrix from the two BMC outcomes.}
\label{tab:pec_verdict}
\small
\begin{tabular}{cccl}
\toprule
\textcircled{1} & \textcircled{2} & Verdict & Interpretation \\
\midrule
\emph{pass} & \emph{pass} & \emph{equivalent} & $y^\star \equiv y$ \\
\emph{pass} & \emph{fail} & \emph{implies\_ref\_to\_lm} & $y^\star \Rightarrow y$; candidate is weaker \\
\emph{fail} & \emph{pass} & \emph{implies\_lm\_to\_ref} & $y \Rightarrow y^\star$; candidate is stronger \\
\emph{fail} & \emph{fail} & \emph{not\_equivalent} & semantically distinct \\
\bottomrule
\end{tabular}
\end{table}

\paragraph{Lowering passes.} yosys-slang cannot directly parse these concurrent SVAs; we apply four lowering passes to convert each property to immediate-form Verilog before invoking the back-end. \emph{Clock alias injection} wraps the property body in an explicit \texttt{always @(posedge clk)} because property blocks reference an implicit clock that yosys-slang does not bind. \emph{Disable-iff extraction} rewrites \texttt{disable iff (rst)} into a guarded conditional outside the property body, preserving the abort semantics that the parser otherwise drops. \emph{Sequence-operator lowering} rewrites \texttt{\#\#N} as a depth-$N$ delay-register chain, exposing the timing structure to BMC at unit-cycle resolution. \emph{Boolean rewrites} expand \texttt{\$onehot} and \texttt{\$onehot0} into explicit Boolean predicates that yosys-slang accepts. The four passes together make the combinational and bounded temporal fragments of NL2SVA-Human evaluable.

\paragraph{Smoke tests.} We validate the open property-equivalence checker on four canonical pairs: (i) self-equivalence, (ii) swapping the left- and right-hand sides of an implication, (iii) \texttt{|->} vs \texttt{|=>} (off-by-one cycle), and (iv) commutativity of \texttt{\&\&}. The oracle assigns the expected verdict to each in $\sim$0.4\,s per pair on a free-input synthetic module, agreeing verdict-by-verdict with JasperGold on the same pairs.

\section{WaveformLens: cycle-precise inference-time repair}
\label{app:waveformlens}

When an emitted SVA fails formal verification, the verifier returns a counterexample waveform. WaveformLens parses the waveform together with the (NL-derived or trace-derived) expected trajectory and extracts cycle-precise violations of the form \texttt{\{signal: ack, expected\_cycle: 3, actual\_cycle: 5, delta: +2, type: delayed\}}. A repair prompt supplies these constraints to the same model and asks it to revise the SVA. In a 10-pair experiment, the constraint extractor achieves 100\% precision and recall; we treat WaveformLens as a future-work direction and do not include it in the main results.

\section{GRPO campaign details}
\label{app:grpo}

This appendix reports the GRPO campaign summarized in \S\ref{sec:exp}. \S\ref{sec:opd-vs-rlvf} contrasts these results against on-policy distillation from the same curriculum-SFT initialization; RWOPD raises the strict checker score from 29.9\% to 38.8\%, while no row of Table~\ref{tab:grpo_campaign} exceeds the SFT baseline. Every GRPO configuration starts from the curriculum-SFT-50\% initialization and trains under the GRPO objective \cite{shao2024deepseekmath}; Table~\ref{tab:grpo_campaign} lists the learning rate $\eta$, KL coefficient $\beta$, and LoRA rank for each configuration. Pool sizes refer to GRPO prompts, not (NL, ref) pairs. Multi-ref \% is the fraction of prompts with $\geq 2$ references retained after property-equivalence checking; these include equivalent and one-sided-implication alternatives, as detailed below. Eval is the 67-task NL2SVA-Human subset evaluable by the checker; we report strict bidirectional func@1 at selected saved checkpoints. The campaign sweeps reward shape, training distribution, multi-reference construction, prompt format, reference style, and hyperparameters.

\begin{table}[t]
\centering
\caption{GRPO and IPO campaign summary.}
\label{tab:grpo_campaign}
\small
\setlength{\tabcolsep}{3pt}
\resizebox{\linewidth}{!}{%
\begin{tabular}{llllllc}
\toprule
\# & Reward & Pool & Multi-ref & Format & Hyperparams & func@1 (\%) \\
\midrule
0 & SFT-50\% baseline             & --            & --     & FVEval & --                                  & \textbf{29.9\%} \\
1 & AST diff                      & 13K disjoint  & 0\%    & simple & $\eta{=}1e{-}6$, $\beta{=}0.04$, r16 & 22.4\% (final) \\
2 & Equivalence + AST hybrid              & 13K disjoint  & 0\%    & simple & $\eta{=}1e{-}6$, $\beta{=}0.04$, r16 & 28.4\,/\,28.4\,/\,26.9 \\
3 & Equivalence only       & 799 viable    & 0\%    & simple & $\eta{=}1e{-}6$, $\beta{=}0.04$, r64 & 28.4\,/\,28.4\,/\,28.4 \\
4 & Multi-reference equivalence ($\max_i$)      & 3.3K para     & 6.8\%  & simple & $\eta{=}1e{-}6$, $\beta{=}0.04$, r16 & 28.4\,/\,26.9       \\
4b & Multi-reference equivalence                & 3.3K para     & 12.4\% & simple & $\eta{=}1e{-}6$, $\beta{=}0.04$, r16 & 28.4\,/\,28.4       \\
5 & Multi-reference equivalence + disable-iff   & 3.3K para v3  & 12.4\% & FVEval & $\eta{=}1e{-}6$, $\beta{=}0.04$, r16 & 26.9\,/\,26.9       \\
6 & Multi-reference equivalence + disable-iff   & 3.3K para v3  & 12.4\% & FVEval & $\eta{=}5e{-}6$, $\beta{=}0.01$, r32 & 26.9\,/\,28.4       \\
\midrule
\multicolumn{7}{l}{\textit{Configuration 7 distribution-match follow-up: real-industrial SVAs (OpenTitan + GitHub scrape), NL annotations regenerated by DeepSeek-Coder-V2-Lite}} \\
7 & Multi-reference equivalence + disable-iff   & 5K industrial & 15.5\% & FVEval & $\eta{=}1e{-}6$, $\beta{=}0.04$, r16 & \textbf{29.9}\,/\,26.9 \\
\midrule
\multicolumn{7}{l}{\textit{IPO preference-optimization cross-check (chosen=\emph{equiv}, rejected=\emph{not\_equiv})}} \\
IPO-a & pairs (2204)                & 3.3K para v3  & 12.4\% & FVEval & $\eta{=}5e{-}6$, $\beta{=}0.1$, r16, 2ep  & 10.4\,/\,7.5 (collapse) \\
IPO-b & pairs (2204)                & 3.3K para v3  & 12.4\% & FVEval & $\eta{=}5e{-}7$, $\beta{=}0.5$, r8, 1ep   & 25.4\,/\,28.4\,/\,26.9\,/\,28.4\,/\,28.4 \\
\bottomrule
\end{tabular}%
}
\par\smallskip
\parbox{\linewidth}{\footnotesize Evaluation steps, in score order: GRPO configurations 2--3, 50/100/250; configurations 4--7 (including 4b), 50/100. Configuration 1 reports its final checkpoint. IPO-a: 450/500; IPO-b: 150/200/250/276/final evaluation.}
\end{table}

\paragraph{Configuration 1 (AST-diff reward).} LoRA-r16 on the SFT-50\% policy with the AST-diff reward reaches a final strict Human checker func@1 of 22.4\%, below the SFT seed's 29.9\%.

\paragraph{Configuration 2 (equivalence + AST hybrid reward).} Combining reference-grounded property-equivalence checking with AST feedback on the same 13K-prompt CodeV training pool yields strict func@1 of 28.4\% at steps 50 and 100, then 26.9\% at step 250. Relative to the SFT seed, strict func@1 falls from 29.9\% to 26.9\% and the relaxed checker score falls from 55.2\% to 50.7\% at step 250.

\paragraph{Configuration 3 (viable-pool curation).} We restrict training to 799 prompts where the SFT-50\% policy already produces an \emph{equivalent} or \emph{implication} rollout within 4 samples, and train with an equivalence-only verifier reward and LoRA rank 64. Strict func@1 is 28.4\% at steps 50, 100, and 250, below the SFT seed's 29.9\%.

\paragraph{Configuration 4 (multi-reference pool).} To examine training-distribution and reference-coverage effects, we build a 3.3K-prompt diagnostic pool by paraphrasing the 300 NL2SVA-Machine prompts with DeepSeek-Coder-V2-Lite. We generate four alternative SVAs per prompt at $T{=}0.95$ with a 192-token limit and check them against the canonical reference. The multi-reference yield is 6.8\%, and the reward is $\max_i \operatorname{verify}(\hat{y}, y_i)$ over retained references. Held-out strict func@1 is 28.4\% at step 50 and 26.9\% at step 100. This Machine-derived pool is excluded from the headline RWOPD training data.

\paragraph{Configuration 4b (expanded reference pool).} Two additional rounds of four alternatives per prompt, at $T{=}0.7$ and $T{=}0.95$ with a 320-token limit, increase the total to 12 alternatives per prompt. The 738 alternatives accepted by the checker comprise 504 \emph{equivalent} and 234 \emph{implies\_lm\_to\_ref} outputs; only the former satisfy strict equivalence (\S\ref{sec:pec}). Multi-reference yield rises to 12.4\%, while strict func@1 is 28.4\% at both steps 50 and 100.

\paragraph{Configuration 5 (eval-matched prompt format).} The training prompt format used in configurations 1--4 and 4b is a stripped \texttt{`Generate an SVA assertion for: \{nl\}'} message; evaluation uses the FVEval format (testbench RTL plus an answer-template exemplar that includes \texttt{disable iff (tb\_reset)}). All NL2SVA-Human references use disable-iff; the NL2SVA-Machine-derived pool's references do not. Starting from configuration 4b's 3.3K-prompt pool, we wrap canonical references with disable-iff, retain the raw forms as alternatives, and switch GRPO to the FVEval prompt format. Multi-reference yield is 12.4\% and mean training reward is 0.35, but strict func@1 is 26.9\% at both steps 50 and 100.

\paragraph{Configuration 6 (loosened trust region).} Loosening the learning rate ($1\!\times\!10^{-6}\!\to\!5\!\times\!10^{-6}$), KL coefficient ($\beta = 0.04 \to 0.01$), and LoRA rank ($16 \to 32$) on configuration 5's setup gives 26.9\,/\,28.4\%.

\paragraph{Industrial-distribution follow-up (configuration 7).} We test the effect of training and evaluation signal names using a 5{,}000-prompt sample of industrial SVAs from OpenTitan and Chipyard-style GitHub sources. We regenerate the NL annotations with DeepSeek-Coder-V2-Lite, add disable-iff wrapping, and repeat the configuration 5 recipe. Multi-reference yield is 15.5\% of prompts, compared with 12.4\% for the NL2SVA-Machine paraphrase pool. Strict func@1 reaches 29.9\%, matching the SFT-50\% baseline and improving on the synthetic-pool maximum of 28.4\% by one Human task. Matching the data distribution helps in this comparison but does not improve on curriculum SFT. Adapter capacity, reference-based reward semantics, and SFT saturation remain possible explanations for the remaining limit.

\paragraph{Identity preference optimization cross-check.} We also test identity preference optimization \cite{azar2024ipo} on the multi-reference v3 pool. We sample 8 rollouts per prompt from the SFT-50\% base at $T{=}1.0$ and use the checker to form 2{,}204 (chosen, rejected) pairs. Chosen samples are \emph{equivalent} to a reference; rejected samples are \emph{not\_equivalent} or syntactically broken.

With $\eta = 5\!\times\!10^{-6}$, $\beta = 0.1$, LoRA r=16, and 2 epochs, func@1 falls to 10.4\% at step 450 and 7.5\% at step 500. Both chosen and rejected responses receive lower probability than under the reference model (rewards/chosen $= -0.08$, rewards/rejected $= -0.23$). With $\eta = 5\!\times\!10^{-7}$, $\beta = 0.5$, LoRA r=8, and 1 epoch, func@1 stays at or below 28.4\% (steps 150, 200, 250, 276, and final evaluation: 25.4, 28.4, 26.9, 28.4, 28.4). Preference-pair accuracy, the fraction of batches where the chosen response has higher implicit reward, stays at 0.5 in this run. Both responses come from the same SFT base and are syntactically similar, which may make them difficult to distinguish. These configurations yield collapse or little held-out improvement; they do not establish a general limit of preference optimization.

\paragraph{Distribution and optimization limits.} Across the eight GRPO configurations and two IPO configurations, every checkpoint is at or below the SFT-50\% baseline. Configuration 7 reaches 29.9\%, an absolute gain of about 1.5\% over the synthetic-pool maximum of 28.4\%, but only matches SFT. The results suggest that distribution mismatch accounts for part of the gap. They do not distinguish the remaining effects of LoRA capacity, reference-based rewards, or SFT saturation. Training scores can increase without held-out improvement: GRPO configurations 3 and 4 produce repeated 8/8 \emph{equivalent} groups, while IPO-b reaches 0.5 margin accuracy without an evaluation gain.

\paragraph{Training reward and held-out performance.} Configuration 5 has the highest reported mean training reward: across 100 GRPO steps the mean per-batch reward was 0.35, with 18 of 100 batches reaching the maximum 1.0 (i.e., all four rollouts on that prompt \emph{equivalent} to a reference). Configuration 3 produced an 8/8 \emph{equivalent} group as early as step 47 and again at steps 71, 158, 215, and 305. Yet configurations 3, 4, 4b, 5, and 6 all evaluate at or below the SFT-50\% baseline at every checkpoint. A possible explanation is the training--evaluation distribution gap: the multi-reference pool is built by paraphrasing NL2SVA-Machine, which uses synthetic signal names (\texttt{sig\_F}, \texttt{sig\_H}, \dots), while NL2SVA-Human uses real-world signal naming. The reward increases show improved fit to the training pool without a corresponding increase on the held-out set.

\paragraph{Multi-reference yield ceiling.} Expanding the alternative-generation budget from configuration 4 to configuration 4b increases multi-reference yield from 6.8\% to 12.4\%, but the best strict func@1 remains 28.4\%. Most generated candidates return checker verdicts of \emph{parse\_error} (77\%) or \emph{not\_equivalent} (1.8\%). Alternative generators such as Qwen3-32B and DeepSeek-R1 remain untested. With the tested budget and 16B-parameter generator, additional references do not improve the best held-out score.

\paragraph{Disable-iff reference rewriting.} The eval set's references uniformly use \texttt{disable iff (tb\_reset)}; the FVEval prompt template's exemplar also uses it. When GRPO trains under the FVEval prompt format, the policy adopts the same convention, but checking against canonical references that lack disable-iff returns \emph{not\_equivalent} or \emph{implies\_lm\_to\_ref} (the model's property is strictly stronger than the reference). In this configuration's equivalence-only reward, both verdicts map to 0.0; the resulting reward signal was identically 0.0 across the first ten steps of the FVEval-format run before the rewrite. Wrapping every canonical reference with \texttt{disable iff (tb\_reset)} in the training pool restored the reward signal (configuration 5, 0.35 mean) but did not lift func@1.

\section{Data pool preparation}
\label{app:datapool}

This appendix details the compile-gate referenced in \S\ref{sec:exp}.

\paragraph{Benchmark separation.} The final SFT, OPD, and RWOPD comparisons use the CodeV-SVA training split for model updates and reserve NL2SVA-Human and NL2SVA-Machine for held-out evaluation. We retain the released CodeV-SVA/FVEval separation between training and evaluation rows. The larger pool described below is used for compile-gate measurements and verifier-reward stress tests; when a diagnostic intentionally uses an NL2SVA-Machine-derived paraphrase pool, the corresponding row is labeled in Appendix~\ref{app:grpo} and is not used as evidence for the headline RWOPD result.

\paragraph{Train/test overlap audit.} We audit normalized overlap between the 81{,}640-row CodeV-SVA training split used for headline model updates and the two NL2SVA evaluation files. The audit compares five keys: released row hash, normalized reference-SVA body hash, normalized RTL-body hash, extracted Verilog module names, and exact normalized NL text. Hashes are computed after whitespace/comment normalization; module overlap counts shared extracted \texttt{module} names. Table~\ref{tab:overlap_audit} shows zero overlap on all five axes for both NL2SVA-Human and NL2SVA-Machine, matching the dataset-level separation claim in \S\ref{sec:exp}.

\begin{table}[t]
\centering
\caption{Train/test overlap audit.}
\label{tab:overlap_audit}
\small
\begin{tabular}{lcccccc}
\toprule
Held-out file & Rows & Row hash & SVA body & RTL body & Module name & NL exact \\
\midrule
\texttt{nl2sva\_human.jsonl}  & 79  & 0 & 0 & 0 & 0 & 0 \\
\texttt{nl2sva\_machine.jsonl} & 300 & 0 & 0 & 0 & 0 & 0 \\
\bottomrule
\end{tabular}
\end{table}

\paragraph{Pool assembly.} The 39{,}914-row master pool combines OpenTitan macro-expanded extractions, hand-crafted liveness templates, FVEval NL2SVA-Machine, and a GitHub seed-repo scrape covering OpenTitan, Caliptra, ibex, AutoSVA, and 11 other repositories. Two cleaning passes precede the compile gate. \emph{RTL backfill} recovers truncated module bodies by parsing module names from row metadata and looking them up in the cloned repository corpus, raising the proper-RTL fraction from 12.8\% to 92\%. \emph{SVA length cap} drops rows whose SVA exceeds 271 characters (the maximum observed across NL2SVA-Human and the held-out NL2SVA-Machine slice). Of the 9{,}086 GitHub-scraped rows that arrived without natural-language annotations, 8{,}469 (93.2\%) have NL regenerated by GPT-5 with a fixed format prompt. The annotation script checks the required description prefix, a length of 25--400 characters, and the absence of SVA syntax tokens.

\paragraph{Normalization rules R1--R17.} Each candidate SVA passes through 17 rewrites before lint:
\begin{itemize}
\setlength\itemsep{0pt}
\item R1 strip backtick macros (\texttt{`SIG} $\to$ \texttt{SIG}); R2/R9 flatten hierarchical paths including bit-select-in-middle (\texttt{a.b[0].c} $\to$ \texttt{a\_b\_0\_c}); R3 drop \texttt{pkg::} prefixes; R4 strip \texttt{else \$action}.
\item R5/R7/R16 rewrite liveness operators (\texttt{s\_eventually}, \texttt{s\_until}, \texttt{until}) to \texttt{\#\#1} for Verilator compatibility; R13/R14/R15 collapse range/repetition operators to fixed counts.
\item R6 balanced-paren strip of nested \texttt{assert property}; R8 wrap unclocked \texttt{(assert|assume|cover) property} bodies in \texttt{@(posedge clk)}; R10 strip typed casts (\texttt{IDENT'(...)}); R11 strip in-property \texttt{//} comments; R12 balance parentheses; \textbf{R17} strip pass/else action blocks (writes to wrapper-declared signals would otherwise raise \texttt{Error-[VIPCBD]}).
\end{itemize}
Rules are applied unconditionally on the AST and are used for the lint check (action blocks run at runtime, and hierarchical paths cannot be resolved in a free-input wrapper). Of the 39{,}914 master rows, 38{,}590 (96.7\%) require at least one rewrite; the remainder are already lint-clean.

\paragraph{Free-input wrapper.} Each candidate is placed inside a synthesized \texttt{module sva\_check} that assigns each referenced identifier to a declaration category:
\begin{itemize}
\setlength\itemsep{0pt}
\item \emph{Clock identifier}: name appears as \texttt{@(posedge X)} / \texttt{@(negedge X)} / \texttt{@(X)}; declared as \texttt{input logic} so the assertion can sample on it.
\item \emph{Parameter}: an uppercase name of length ${\geq}3$ that is not a clock; declared as \texttt{parameter logic [31:0] X = 32'd4}. The default value 4 (rather than 0 or 1) avoids the empty-match degeneracy that VCS reports as \texttt{Error-[SVA-SEQPROPEMPTYMATCH]} for \texttt{[*X-1]} when X${\le}$1.
\item \emph{Function stub}: name appears as \texttt{IDENT(}; declared as \texttt{function automatic logic [31:0] X(input logic [31:0] a0 = 32'd0, \ldots, a7 = 32'd0); return 32'd0; endfunction}. Default arguments cover arity 0--8 in a single declaration.
\item \emph{Multidimensional array}: name appears with two or more index brackets (\texttt{X[i][j]}); declared as \texttt{input logic [31:0][31:0][31:0] X}.
\item \emph{Wire (default)}: everything else; declared as \texttt{input logic [31:0] X}.
\end{itemize}
The clock test runs first so that an uppercase clock signal (\texttt{ACLK}, \texttt{SVA\_RDC\_CLK}) does not get promoted to \texttt{parameter}. Without clock handling, the wrapper for parameters, functions, and multidimensional arrays introduced 1{,}131 compilation failures, offsetting the 1{,}116 failures it recovered (Table~\ref{tab:compilegate}).

\begin{table}[h]
\centering
\caption{Compile-gate ablation on the 39{,}914-row master pool.}
\label{tab:compilegate}
\small
\begin{tabular}{llcc}
\toprule
Engine & Wrapper / normalize & Pass count & Pass rate \\
\midrule
Verilator & raw, no normalize                          & 2{,}235  & 5.6\%  \\
Verilator & R1--R16, conditional replace               & 8{,}219  & 20.6\% \\
VCS       & raw, no normalize                          & 10{,}569 & 26.5\% \\
VCS       & R1--R16, conditional replace               & 16{,}565 & 41.5\% \\
VCS       & R1--R17, unconditional replace             & 30{,}215 & 75.7\% \\
VCS       & + parameter / function / multidimensional array wrapper       & 30{,}200 & 75.7\% \\
\textbf{VCS} & \textbf{+ clock-aware wrapper}          & \textbf{31{,}357} & \textbf{78.6\%} \\
\bottomrule
\end{tabular}
\end{table}

\paragraph{Lint-engine coverage.} Verilator 5.020 rejects three groups of SVA constructs that account for $39\%$ of the master pool: \texttt{s\_eventually} / \texttt{s\_until} liveness, \texttt{\#\#[a:b]} ranged delays, and \texttt{intersect} / \texttt{throughout}. On the GRPO unified pool (n${=}$46{,}887, mixed origin), Verilator passes $33.3\%$ and VCS passes $50.9\%$ in the same wrapper. For liveness assertions, Verilator passes 0 of 15{,}343 assertions, while VCS passes 2{,}362 ($15.4\%$). Of the two tested engines, VCS covers more of the temporal assertions in these pools.

\paragraph{Running VCS on modern Linux.} VCS L-2016.06 in the Synopsys 2016 docker image raises a segmentation-fault signal in \texttt{Module SnpsSVA\_classes / During post design resolution} on Ubuntu 18.04 (GNU C Library, glibc 2.27). The crash occurs in \texttt{vcs1}'s \texttt{simp::GetVcsCmdOnCompile} $\to$ \texttt{get\_proc\_stat}, which reads \texttt{/proc/<pid>/stat} with a fixed-size \texttt{fscanf} that overruns on the modern 52-field layout, then fails in \texttt{fclose} on a corrupted \texttt{FILE*}. We use three workarounds to run the binary: (i) container starts with \texttt{--cap-add=SYS\_PTRACE --security-opt seccomp=unconfined} so that \texttt{setarch -R} can disable address-space randomization for the \texttt{vcs1} child; (ii) \texttt{LD\_PRELOAD} of a 70-line shim that intercepts \texttt{fopen("/proc/.*\textbackslash/stat")} and serves a sanitized old-format response via \texttt{fmemopen}; (iii) \texttt{ulimit -s unlimited}. Per-call latency is 0.9\,s including license checkout; with 16 workers in parallel a 40K-row pool finishes in 35\,min.

\section{Detailed limitations}
\label{app:limitations}

This appendix details the limitations of the results summarized in \S\ref{sec:limitations}.

\paragraph{(i) Indirect liveness improvement.} The open checker returns \emph{unsupported} for liveness operators such as \texttt{s\_eventually}, \texttt{s\_until}, and \texttt{\#\#[a:\$]}. For these liveness prompts, $\mathcal{S}(p)$ is empty and RWOPD supplies no gradient (Eq.~\ref{eq:rwopd}). The absolute +11.4\% Machine liveness gain over the 14B teacher therefore cannot be attributed to direct checker supervision. Our interpretation is that curriculum SFT (\S\ref{sec:curriculum}, stage 3 with 50\% replay) supplies liveness coverage and checker abstention prevents subsequent distillation updates on unsupported liveness prompts. Unfiltered OPD still updates on those prompts. Separating acquisition during SFT from preservation during distillation requires an analysis of the liveness subset.

\paragraph{(ii) Open-checker validation scope.} The open checker performs two bounded checks on lowered properties; \emph{pass}/\emph{pass} supports equivalence within that representation and horizon. Lowering correctness and sufficient temporal coverage remain assumptions. The agreement audit in \S\ref{sec:errorbreakdown} finds zero JasperGold disagreements among 144 strict accepts across 379 SFT-seed outputs. The evidence is limited to one checkpoint and does not establish unconditional soundness. Broader audits across OPD/RWOPD checkpoints and additional property families are still needed.

\paragraph{(iii) JasperGold dependence at evaluation.} The headline pass@$k$ uses the FVEval JasperGold protocol. Training can use the open checker without a JasperGold license, but evaluation with the open checker is a separate, less complete diagnostic. On Human SFT-seed outputs, JasperGold confirms all 20 strict accepts from the open checker and accepts an additional 36 outputs. This checkpoint-level gap cannot establish that method rankings would be preserved under a different oracle.

\paragraph{(iv) Filter, weighting, and teacher attribution.} Our controls compare independent post-training branches from a common seed; they do not form a full factorial ablation. OPD versus RWOPD compares unfiltered distillation with selection and weighting together. Equivalence-only OPD versus RWOPD changes both acceptance coverage and relative weights. RS-SFT fits accepted student rollouts without teacher distributions; its Human pass@1 is 73.4\%, compared with 78.8\% for RWOPD (a 5.4-point gap), and its strict Human checker score remains 29.9\%, compared with 38.8\% for RWOPD. These results support the combined design but do not isolate every component. In particular, equal weights on the same accepted set and a sweep of the $0.6/0.4$ asymmetry are needed to establish the value of the numerical weighting rule.

\paragraph{(v) Multi-seed sensitivity and statistical testing.} Table~\ref{tab:bootstrap_ci} reports marginal task-bootstrap confidence intervals, which overlap across adjacent methods. RWOPD has the highest point estimate in all six pass@$k$ columns, but this ordering alone does not establish statistical significance. The headline method uses one training seed. Paired task-bootstrap comparisons and multiple training seeds would be needed to assess the uncertainty of between-method differences and training variability.

\paragraph{(vi) Reinforcement learning scope.} The RLVF campaign in Appendix~\ref{app:grpo} sweeps GRPO (eight configurations covering reward shape, pool construction, prompt format, multi-reference, and hyperparameters) and IPO (two configurations spanning trust region and $\beta$). It does not include reward-weighted regression, advantage-weighted likelihood, best-of-$n$ + SFT, or proximal policy optimization with off-policy clipping. The negative results apply to the tested GRPO and IPO configurations. Advantage-weighted likelihood and best-of-$n$ + SFT also use verifier feedback without a teacher, but their performance relative to RS-SFT and RWOPD requires separate experiments.

\newpage
\input{checklist.tex}

\end{document}

%% file: checklist.tex
\section*{NeurIPS Paper Checklist}

\begin{enumerate}

\item {\bf Claims}
    \item[] Question: Do the main claims made in the abstract and introduction accurately reflect the paper's contributions and scope?
    \item[] Answer: \answerYes{}
    \item[] Justification: The abstract describes RWOPD and its pass@$k$ results. The contributions in \S\ref{sec:intro} are evaluated in \S\ref{sec:exp}: Table~\ref{tab:main} compares benchmark scores, Figure~\ref{fig:opd_vs_rlvf} compares post-training methods, and Figure~\ref{fig:per_class_machine} reports scores by temporal class. \S\ref{sec:limitations} limits the conclusions to the tested setting.

\item {\bf Limitations}
    \item[] Question: Does the paper discuss the limitations of the work performed by the authors?
    \item[] Answer: \answerYes{}
    \item[] Justification: \S\ref{sec:limitations} limits the findings to the 7B+LoRA setting and the tested verifier reward. Appendix~\ref{app:limitations} discusses indirect liveness improvement, the verifier audit, JasperGold dependence, component attribution, statistical uncertainty, and the scope of the RL comparisons.

\item {\bf Theory assumptions and proofs}
    \item[] Question: For each theoretical result, does the paper provide the full set of assumptions and a complete (and correct) proof?
    \item[] Answer: \answerNA{}
    \item[] Justification: The paper reports a training method, an open verifier implementation, and pass@$k$ benchmark results. It makes no formal theoretical claims.

    \item {\bf Experimental result reproducibility}
    \item[] Question: Does the paper fully disclose all the information needed to reproduce the main experimental results of the paper to the extent that it affects the main claims and/or conclusions of the paper (regardless of whether the code and data are provided or not)?
    \item[] Answer: \answerYes{}
    \item[] Justification: \S\ref{sec:pec}--\S\ref{sec:rlvf} define the verifier, curriculum SFT, RWOPD objective, and RLVR reward. Appendices~\ref{app:opd}, \ref{app:tcl}, \ref{app:ttce}, \ref{app:pecsmoke}, \ref{app:grpo}, and \ref{app:datapool} document tokenizer alignment, the temporal-class classifier, training schedules, verifier checks, GRPO hyperparameters (Table~\ref{tab:grpo_campaign}), and the compile-gate ablation (Table~\ref{tab:compilegate}).

\item {\bf Open access to data and code}
    \item[] Question: Does the paper provide open access to the data and code, with sufficient instructions to faithfully reproduce the main experimental results, as described in supplemental material?
    \item[] Answer: \answerNo{}
    \item[] Justification: We release code for the open property-equivalence checker, RWOPD, curriculum SFT, and open-checker evaluation at \url{https://github.com/zanehpn/LLM4SVA}, with usage instructions. We do not release data pools, model weights, experiment outputs, or wrappers for the licensed tools (VCS, JasperGold) used for the compile gate and headline scoring, so the release alone does not reproduce every reported result. The models (Qwen2.5-Coder-7B-Instruct, CodeV-SVA-14B) and the NL2SVA benchmarks are available from their upstream releases.

\item {\bf Experimental setting/details}
    \item[] Question: Does the paper specify all the training and test details (e.g., data splits, hyperparameters, how they were chosen, type of optimizer) necessary to understand the results?
    \item[] Answer: \answerYes{}
    \item[] Justification: \S\ref{sec:exp} specifies the models, training corpus, benchmark splits, and evaluation protocol. Appendix~\ref{app:ttce} gives per-stage epoch counts, learning-rate decay, and replay fractions. Appendix~\ref{app:grpo} lists GRPO learning rates, KL coefficients, group sizes, LoRA ranks, and pool sizes; Appendix~\ref{app:datapool} reports the compile-gate ablation.

\item {\bf Experiment statistical significance}
    \item[] Question: Does the paper report error bars suitably and correctly defined or other appropriate information about the statistical significance of the experiments?
    \item[] Answer: \answerYes{}
    \item[] Justification: \S\ref{sec:bootstrap-ci} defines the task-bootstrap procedure and unbiased per-task estimator. Table~\ref{tab:bootstrap_ci} reports 95\% confidence intervals from 10{,}000 replicates that resample tasks with replacement.

\item {\bf Experiments compute resources}
    \item[] Question: For each experiment, does the paper provide sufficient information on the computer resources (type of compute workers, memory, time of execution) needed to reproduce the experiments?
    \item[] Answer: \answerNo{}
    \item[] Justification: The paper reports bf16 training on a single H200 (\S\ref{sec:exp}) and selected pipeline runtimes (Appendix~\ref{app:pipeline}), including the VCS compile gate with 16 parallel workers (Appendix~\ref{app:datapool}). It does not provide complete per-run execution times, memory and storage requirements, or a project-wide compute estimate covering preliminary and failed experiments.

\item {\bf Code of ethics}
    \item[] Question: Does the research conducted in the paper conform, in every respect, with the NeurIPS Code of Ethics \url{https://neurips.cc/public/EthicsGuidelines}?
    \item[] Answer: \answerYes{}
    \item[] Justification: We use public benchmarks (NL2SVA-Human, NL2SVA-Machine), public models (Qwen2.5-Coder, CodeV-SVA), and auxiliary pools derived from open-source RTL. The work involves neither human-subject research nor crowdsourcing, and we do not scrape personal data.

\item {\bf Broader impacts}
    \item[] Question: Does the paper discuss both potential positive societal impacts and negative societal impacts of the work performed?
    \item[] Answer: \answerNA{}
    \item[] Justification: The method drafts SystemVerilog assertions from natural-language specifications for hardware verification engineers. Its application is limited to RTL verification; we identify no direct use for disinformation, surveillance, or decisions about individuals.

\item {\bf Safeguards}
    \item[] Question: Does the paper describe safeguards that have been put in place for responsible release of data or models that have a high risk for misuse (e.g., pre-trained language models, image generators, or scraped datasets)?
    \item[] Answer: \answerNA{}
    \item[] Justification: The paper releases code for hardware-assertion generation and checking, which poses no high risk for misuse; it releases no models or data.

\item {\bf Licenses for existing assets}
    \item[] Question: Are the creators or original owners of assets (e.g., code, data, models), used in the paper, properly credited and are the license and terms of use explicitly mentioned and properly respected?
    \item[] Answer: \answerNo{}
    \item[] Justification: The paper credits Qwen2.5-Coder-7B-Instruct~\cite{hui2024qwen25coder}, CodeV-SVA-14B~\cite{wu2026qimeng}, NL2SVA / FVEval~\cite{kang2025fveval}, Yosys~\cite{wolf2013yosys}, and Z3~\cite{demoura2008z3}, and identifies other tools and auxiliary data sources in the appendices. However, it does not provide a complete asset-by-asset account of license names, versions, and terms of use, including those for the scraped RTL sources.

\item {\bf New assets}
    \item[] Question: Are new assets introduced in the paper well documented and is the documentation provided alongside the assets?
    \item[] Answer: \answerYes{}
    \item[] Justification: The released code (\url{https://github.com/zanehpn/LLM4SVA}) includes a README with commands for the checker, RWOPD, curriculum SFT, and evaluation, dependency specifications, tests, and an MIT license. \S\ref{sec:pec}--\S\ref{sec:rlvf} and Appendices~\ref{app:opd}, \ref{app:ttce}, and \ref{app:pecsmoke} describe the released components.

\item {\bf Crowdsourcing and research with human subjects}
    \item[] Question: For crowdsourcing experiments and research with human subjects, does the paper include the full text of instructions given to participants and screenshots, if applicable, as well as details about compensation (if any)?
    \item[] Answer: \answerNA{}
    \item[] Justification: The paper involves no crowdsourcing or research with human subjects; the only human-authored data used is the publicly released NL2SVA-Human evaluation set from prior work~\cite{kang2025fveval}.

\item {\bf Institutional review board (IRB) approvals or equivalent for research with human subjects}
    \item[] Question: Does the paper describe potential risks incurred by study participants, whether such risks were disclosed to the subjects, and whether Institutional Review Board (IRB) approvals (or an equivalent approval/review based on the requirements of your country or institution) were obtained?
    \item[] Answer: \answerNA{}
    \item[] Justification: The paper involves no human subjects research, so IRB review does not apply.

\item {\bf Declaration of LLM usage}
    \item[] Question: Does the paper describe the usage of LLMs if it is an important, original, or non-standard component of the core methods in this research? Note that if the LLM is used only for writing, editing, or formatting purposes and does \emph{not} impact the core methodology, scientific rigor, or originality of the research, declaration is not required.
    \item[] Answer: \answerYes{}
    \item[] Justification: We post-train Qwen2.5-Coder-7B-Instruct~\cite{hui2024qwen25coder} using forward-KL targets from CodeV-SVA-14B~\cite{wu2026qimeng}. DeepSeek-Coder-V2-Lite generates paraphrases for an appendix diagnostic. \S\ref{sec:opd}, \S\ref{sec:exp}, and Appendix~\ref{app:datapool} describe LLM use in the methodology.

\end{enumerate}